\documentclass[a4paper,10pt,twoside]{cpc-hepnp}

\usepackage{multicol}
\usepackage{graphicx}
\usepackage{booktabs}
\usepackage{amssymb,bm,mathrsfs,bbm,amscd}
\usepackage[tbtags]{amsmath}
\usepackage{lastpage}
\usepackage{CJK}
\usepackage{epstopdf}

\begin{document}

\fancyhead[c]{\small Chinese Physics C~~~Vol. xx, No. x (202x) xxxxxx}
\fancyfoot[C]{\small 010201-\thepage}

\footnotetext[0]{Received 14 March 2025}

\title{Charged-current quasielastic neutrino scattering off nuclei with nucleon-nucleon short-range correlations\thanks{supported by the National Natural Science Foundation of China (Grants No. 12475135 and No. 12035011), by the Shandong Provincial Natural Science Foundation, China (Grant No. ZR2020MA096), and by the Fundamental Research Funds for the Central Universities (Grant No. 22CX03017A).}}

\author{%
      Jian Liu $^{1;1)}$\email{liujian@upc.edu.cn}%
\quad Qiang Su $^{1}$
\quad Qinglin Niu $^{1}$
\quad Lei Wang $^{2}$
\quad Zhongzhou Ren $^{2;2)}$\email{zren@tongji.edu.cn}%
}
\maketitle

\address{%
$^1$ College of Science,  China University of Petroleum (East China),
 Qingdao 266580, China\\
$^2$ School of Physics Science and Engineering, Tongji University, Shanghai 200092, People's Republic of China\\
}

\begin{abstract}
In recent years, many studies on neutrino-nucleus scattering have been carried out to investigate nuclear structures and the interactions between neutrinos and nucleons. This paper develops a charged-current quasielastic (CCQE) neutrino-nucleus scattering model to explore the nuclear mean-field dynamics and short-range correlation effects. In this model, the nuclear structure effect is depicted using the scaling function $f(\psi)$, while the neutrino-nucleon interaction is represented by the elementary weak cross section $\sigma_0$. Results indicate that the double-differential cross section of scattered muon is influenced by the energy $E$ and momentum $\mathbf{p}$ of nucleon in nuclei, and the total cross section depends primarily on the incident neutrino energy $E_\nu$. Furthermore, incorporating short-range correlations yields the flux-integrated differential cross sections at high-$T_\mu$ region producing larger values, a longer tail, and achieving better experimental consistency. It eventually elucidates the physical relationship between the neutrino-nucleus scattering cross section and the variation in incident neutrino energy. The studies in this paper furnishes insights for the research of nucleon dynamics and provides detailed examinations of the neutrino-nucleus scattering mechanism.
\end{abstract}

\begin{keyword}
Nucleon-nucleon short-range correlations, Scaling function, Charged-current quasielastic neutrino scattering
\end{keyword}

\begin{pacs}
21.10.Gv, 24.80.+y, 25.30.Bf)
\end{pacs}

\footnotetext[0]{\hspace*{-3mm}\raisebox{0.3ex}{$\scriptstyle\copyright$}2013
Chinese Physical Society and the Institute of High Energy Physics
of the Chinese Academy of Sciences and the Institute
of Modern Physics of the Chinese Academy of Sciences and IOP Publishing Ltd}%

\begin{multicols}{2}

\section{Introduction}

As a crucial tool in particle physics, nuclear physics and cosmology, neutrino scattering plays a pivotal role in understanding fundamental particle interactions\cite{bahcall1987neutrino}, revealing the nucleon-nucleon short-range correlations (\textit{NN}-SRC) \cite{van2016}, and exploring cosmic evolution \cite{akimov2017observation,kim2024}. Over recent decades, experiments on neutrino-nucleus scattering, including those conducted at CE$\nu$NS \cite{ross2024}, NOMAD \cite{lyubushkin2009study}, MiniBooNE \cite{aguilar2010measurement}, MINOS \cite{adamson2015study}, and T2K \cite{ahn2003indications}, have significantly advanced our understanding of nuclear structure. The quasielastic neutrino scattering can be divided into two primary classifications: the charged-current quasielastic (CCQE) neutrino scattering $\nu_l\left(\tilde{\nu}_l\right)+n(p) \rightarrow p(n)+\mu^{-}\left(\mu^{+}\right)$ and the neutral-current quasielastic (NCQE) neutrino scattering $\nu_l\left(\tilde{\nu_l}\right)+n(p) \rightarrow \nu_l\left(\tilde{\nu_l}\right)+n(p)$ \cite{van2018dependence,ankowski2012consistent,ivanov2014charged}. The CCQE scattering is produced by the exchange of $W^{ \pm}$ bosons, enabling a charge transfer at the interaction point. Compared to CCQE scattering, in NCQE neutrino scattering, $Z^0$ bosons play a pivotal role, and no charge exchange occurs \cite{formaggio2012ev,barbaro2018asymmetric}. The CCQE neutrino scattering offers a unique perspective to study the nuclear internal dynamics, due to its charge-changing interaction between neutrinos and nucleons. In scattering process, the incident neutrinos interact with individual nucleons, enabling the study of nucleon interactions \cite{van2004quasielastic}. Therefore, in the domain of CCQE, the changes in the nucleon momentum distribution (NMD) caused by nucleon correlations can be observed more effectively.

The plane-wave impulse approximation (PWIA) is a essential method for studying the neutrino-nucleus scattering. In the framework of PWIA, the cross sections of muon can be expressed as a product of two parts: the elementary weak cross section $\sigma_0$ showing the neutrino scattering off a free nucleon and the scaling function $f(\psi)$ reflecting distributions of nucleons. The scaling function was first constructed based on the relativistic Fermi gas (RFG) model \cite{ivanov2008superscaling}. However, the classical RFG model cannot well reproduce the experimental data. Therefore, researchers have endeavored to develop scaling functions by introducing the nuclear dynamics. It is an effective method to derive scaling function from the sophisticated nuclear structure model \cite{antonov2011scaling}.

In past years, mean-field models based on energy density functionals (EDFs) have been used to investigate the nuclear properties \cite{ring1996relativistic,bender2003self,martinez2024semi}. Resently, the theoretical and experimental studies have revealed the important role of \textit{NN}-SRC, including the emergence of high momentum tails and diminished occupancy of low-lying nuclear states \cite{casale2023center,niu2022effects,hen2014momentum}. Therefore, the combination of mean-field and \textit{NN}-SRC effects is expected to offer the theoretical explanation for neutrino-nucleus scattering. There are multiple approaches to introduce \textit{NN}-SRC effects, and it is a feasible choice using the light-front dynamics (LFD) approach \cite{carbonell1995relativistic}. The LFD method calculates NMD of the correlation part $n_{\text {corr}}(p)$ by empirically rescaling the high-momentum components of NMDs of the deuteron \cite{wang2021nucleon}. Compared to the momentum distribution of protons, examining the momentum distribution of neutrons is more challenging, primarily due to its electrically neutral. The focus of this paper aims to explore the properties of neutrons through the neutrino-neutron reaction $\nu_l+n \rightarrow p+\mu^{-}$. Given that neutrinos interact with neutrons only via weak forces, the CCQE neutrino scattering presents distinct advantages in probing the momentum distribution of neutrons.

The main work of this paper is outlined as follows. Firstly, we investigate the spectral function $S(\mathbf{p}, E)$, which represents the joint probability distribution to find nucleon in the target nucleus with momentum $\mathbf{p}$ and removal energy $E$. In calculating $S(\mathbf{p}, E)$, the mean-field component is calculated from  axially deformed Hartree-Fock-Bogoliubov (HFB) \cite{wang2025} model, while the correlation part is introduced through the LFD method. Subsequently, based on the spectral function $S(\mathbf{p}, E)$, the scaling function $f(\psi)$ is derived to effectively represent the structural information of the target nucleus. Finally, we focus on building the CCQE neutrino-nucleus scattering model, in which the elementary weak cross section $\sigma_0$ is calculated based on the nucleon form factor and the nuclear structure is introduced through the scaling function $f(\psi)$. The influence of \textit{NN}-SRC on the neutron momentum distribution is evaluated using experimental observables.
	
	This paper is organized as follows: In Sec. 2, we construct the scaling function $f(\psi)$, and provide the corresponding formulas for CCQE neutrino-nucleus scattering. In Sec. 3,  the results of neutrino scattering cross sections are presented and discussed. Finally, a summary is given in Sec. 4.

\section{Theoretical framework}

This section is organized into three parts. Firstly, we provide the theoretical formulas for the CCQE neutrino-nucleus scattering cross section. Secondly, the scaling function $f(\psi)$ reflecting nuclear structure information is constructed using the spectral function $S(\mathbf{p}, E)$. Finally, The spectral function $S(\mathbf{p}, E)$ are derived employing the HFB theory and LFD method.

\subsection{CCQE neutrino cross section}

The CCQE neutrino scattering refers to the process $\nu_l+n \rightarrow p+\mu^{-}$ where a neutrino interacts with a target nucleus, resulting in the emission of a single muon. In this paper, we define the energy of the incoming neutrino as $E_\nu$ and the kinetic energy of the outgoing muon as $T_\mu$. The mass of muon is $m_\mu^{\prime}$ and total space scattering angle is expressed as $\Omega^{\prime}$. The momentum transfer is denoted by $q$, and the energy transfer is denoted by $\omega$.

The neutrino double-differential cross section is expressed as the product of the elementary weak scattering cross section $\sigma_0$ and the structure function $\mathcal{F}_{+}^2$
	\begin{equation}\label{1}
			\frac{d^2 \sigma}{d \Omega^{\prime} d T_\mu}=\sigma_0 \mathcal{F}_{+}^2,
	\end{equation}
where $\sigma_0$ represents the scattering cross section for neutrino interactions with a free nucleon
\begin{equation}\label{2}
		\sigma_0=\frac{G^2 \cos ^2 \theta_c}{2 \pi^2} k^{\prime} T_\mu \cos ^2 \frac{\tilde{\theta}}{2} .
\end{equation}
Here $G=1.166 \times 10^{-11} {\rm \,MeV}^{-2}$ describes the strength of the weak interaction, and the Cabibbo angle  $\cos \theta_c=0.975$ preserves the universality of the weak interaction. $k^{\prime}$ is the momentum of outgoing muon. The generalized scattering angle $\tilde{\theta}$ in Eq. (\ref{2}) is \cite{amaro2005using}
\begin{equation}\label{3}
		\tan ^2 \frac{\tilde{\theta}}{2}=\frac{\left|Q^2\right|}{\left(E_\nu+T_\mu\right)^2-q^2},
\end{equation}
with $Q^2=\omega^2-q^2$.

The structure function $\mathcal{F}_{+}^2$ in Eq. (\ref{1}), which contains the neutron momentum distributions, neutron energy distributions, nucleon form factors, and other nuclear structure details. $\mathcal{F}_{+}^2$ can be presented as a generalized Rosenbluth decomposition having charge-charge, charge-longitudinal, longitudinal-longitudinal, and two types of transverse responses \cite{amaro2005using}
\begin{equation}\label{4}
		\mathcal{F}_{+}^2=\widehat{V}_{CC} R_{CC}+2 \widehat{V}_{CL} R_{CL}+\widehat{V}_{LL} R_{LL}+\widehat{V}_T R_T+2 \widehat{V}_{T^{\prime}} R_{T^{\prime}} ,
\end{equation}
where the kinematical factors $\hat{V}_K\left(K=C C, C L, L L, T, T^{\prime}\right)$ come from the leptonic tensor. The response functions $R_K$ in Eq. (\ref{4}) are written as
\begin{equation}\label{13}
		R_K=\mathcal{N} \mathit{\Lambda}_0 U_K f(\psi), \quad K=C C, C L, L L, T, T^{\prime},
\end{equation}
where $\mathcal{N}$ is the neutron number. $\mathit{\Lambda}_0$ and $U_K$ in Eq. (\ref{13}) are the response factor and the single-nucleon responses, which can be found in \cite{amaro2005semirelativistic}. $f(\psi)$ is the $\psi$ scaling function that contains the nuclear structure information. The detailed description is provided in the next subsection.

After obtaining the double-differential cross section ${d^2 \sigma}/{d \Omega^{\prime} d T_\mu}$, we evaluate the flux differential cross section for CCQE process averaged over the neutrino flux $\Phi\left(E_\nu\right)$
\begin{equation}\label{18}
		\frac{d^2 \sigma}{d T_\mu d \cos \theta_\mu}=\frac{1}{\Phi_{\text {tot }}} \int\left[\frac{d^2 \sigma}{d T_\mu d \cos \theta_\mu}\right]_{E_\nu} \Phi\left(E_\nu\right) d E_\nu,
\end{equation}
where $\theta_\mu$ is the scattering angle of the outgoing muon. The neutrino flux $\Phi\left(E_\nu\right)$ represents the probability of neutrinos interacting with other matter at different incident energies and $\Phi_{\text {tot }}$ is the total integrated neutrino flux factor  \cite{aguilar2010measurement,ivanov2015neutral,lu2024}.

By integrating over the scattering angle $\theta_\mu$ and incident neutrino energy $E_\nu$ in the double-differential cross section of Eq. (\ref{1}), we can gain the differential cross section as a function of the outgoing muon kinetic energy $T_\mu$,
\begin{equation}\label{19}
		\left\langle\frac{d \sigma}{d T_\mu}\right\rangle=\frac{1}{\Phi_{\text {tot }}} \int \Phi\left(E_\nu\right) \int\left[\frac{d^2 \sigma}{d T_\mu d \cos \theta_\mu}\right]_{E_\nu} d \cos \theta_\mu d E_\nu.
\end{equation}
Similarly, the differential cross section as a function of $\theta_\mu$ can be obtained by integrating over both the outgoing muon kinetic energy $T_\mu$ and the incident neutrino energy $E_\nu$
\begin{equation}\label{20}
		\left\langle\frac{d \sigma}{d \cos \theta_\mu}\right\rangle=\frac{1}{\Phi_{\text {tot }}} \int \Phi(E_\nu) \int\left[\frac{d^2 \sigma}{d T_\mu d \cos \theta_\mu}\right]_{E_\nu} d T_\mu d E_\nu .
\end{equation}
Finally, the total cross section $\sigma_T$ of neutrino-nucleus scattering is expressed by integrating over $\theta_\mu$ and $T_\mu$ in the double-differential cross section
\begin{equation}\label{21}
		\sigma_T(E_\nu)=\iint\left[\frac{d^2 \sigma}{d T_\mu d \cos \theta_\mu}\right]_{E_\nu} d T_\mu d \cos \theta_\mu \text {. }
\end{equation}

\subsection{$\psi$ scaling function in CCQE cross section}

Scaling method is a powerful tool to study neutrino scattering in quasielastic region, which elucidates the dynamics of neutrino interactions with nucleons inside the nucleus \cite{ivanov2024superscaling}. The $\psi$ scaling function provides critical insights into the target nucleus, capturing the distribution of momentum and energy among its nucleons. By introducing a kinematical variable, the scaling variable $\psi$, which is solely dependent on the momentum transfer $q$ and the energy transfer $\omega$ \cite{antonov2005superscaling}
	\begin{equation}\label{22}
			\psi=\frac{1}{\sqrt{\xi_F}} \frac{\lambda-\tau}{\sqrt{(1+\lambda) \tau+\kappa \sqrt{\tau(1+\tau)}}},
	\end{equation}
	 with $\lambda \equiv \omega / 2 m_n$ and $\tau \equiv\kappa^2-\lambda^2$ \cite{antonov2011scaling}. $m_n$is neutron mass. $\kappa$ and $\xi_F$ are the dimensionless fermi kinetic energy and the transfer momentum factor, respectively, as defined in \cite{amaro2005semirelativistic}. The scaling function can be derived from the structure functions \cite{degli1991scaling}
	\begin{equation}\label{23}
			F(q, \psi)=2 \pi \int_{E_{\min }}^{E_{\max }(q, \psi)} d E \int_{P_{\min }(q, \psi, E)}^{P_{\max }(q, \psi, E)}  d pS(p, E) p,
	\end{equation}
	where $S(\mathbf{p}, E)$ is the neutron spectral function, with a detailed description provided in the Sec. 2.3. Through energy conservation in the scattering process, the upper and lower limits of the energy integration in Eq. (\ref{23}) are
\begin{subequations}
	\begin{gather}
	E_{\min }=M_{A-1}+m_n-M_A, \\ 	
	E_{\max }=M_A^*-M_A, \quad\quad\quad\enspace
	\end{gather}
	\label{24}%
\end{subequations}
where $E_{\min }$ denotes the single-neutron separation energy and $M_A^*$ is the effective mass of the system composed of the residual nucleons. We further introduce the momentum conservation in CCQE process
	\begin{equation}\label{26}
			 \omega+M_{A}=\sqrt{{m_n}^2+(\mathbf{k}+\mathbf{q})^2}+\sqrt{{M}_{A-1}^{* 2}+\mathbf{k}^2},
	\end{equation}
where the angles between $\mathbf{k}$ and $\mathbf{q}$ ranges from $0^{\circ}$ to $180^{\circ}$. By substituting the angles of $0^{\circ}$ and $180^{\circ}$ into Eq. (\ref{26}), we obtain the upper and lower limits of momentum integration in Eq. (\ref{23}). The upper and lower bounds of the momentum integral can be found in our previous work \cite{wang2023new}.

At large $q$, the scaling function depends only on a single kinematic variable $\psi$, and we can obtain the dimensionless scaling function $f(\psi)$ of Eq. (\ref{13}) is \cite{ivanov2014charged}
	\begin{equation}\label{32}
			f(\psi)=F(q, \psi) \times p_F .
	\end{equation}
In Eq. (\ref{32}), $p_F$ denotes the Fermi momentum of the nucleus. Due to the CCQE process, where neutrinos only react with neutrons, the spectral function for $f(\psi)$ calculations only considers the energy and momentum distribution of neutrons.

\subsection{The nuclear spectral function}

The neutron spectral function $S(\mathbf{p}, E)$ in Eq. (\ref{24}) represents the probability of finding a neutron with momentum $\mathbf{p}$ and removal energy $E$ in nuclei \cite{wang2023influence}. In this paper, the calculations of spectral function $S(\mathbf{p}, E)$ are divided into two parts: the mean-field (MF) part $S(\mathbf{p}, E)$ and the correlation component $S_{\text {corr}}(\mathbf{p}, E)$ \cite{niu2022effects}
\begin{equation}\label{33}
		S(\mathbf{p}, E)=S_{\mathrm{MF}}(\mathbf{p}, E)+S_{\text {corr}}(\mathbf{p}, E).
\end{equation}
The MF part $S_{\mathrm{MF}}(\mathbf{p}, E)$ is dominated by the single-neutron properties at low energy and low momentum
\begin{equation}\label{34}
		S_{\mathrm{MF}}(\mathbf{p}, E)=\sum_i C_i n_i(\mathbf{p}) L_i\left(E-E_i\right),
\end{equation}
where $C_i$ is the corresponding occupation number of the single-neutron state $i$. The Lorentzian function $L_i$ describes the finite width in energy dependence and $E_i$ is the eigenvalue of the energy of the state $i$. The detailed parameters of $E_i$ and their values are sourced from Refs. \cite{niu2022effects} and \cite{ivanov2019realistic}. The single-neutron momentum distribution $n_i(\mathbf{p})$ are obtained by applying the Fourier transform to the single-particle Hartree-Fock nucleon wave function from $r$-space to $p$-space \cite{wang2021nucleon}. The wave function is calculated using HFB2.0 code \cite{stoitsov2013axially}, which allows for axially symmetric deformations.

The correlation component $S_{\text {corr}}(\mathbf{p}, E)$ in Eq. (\ref{33}) can be obtained from
\begin{equation}\label{35}
		S_{\text {corr}}(\mathbf{p}, E)=n_{\text {corr}}(\mathbf{p}) \frac{m_n}{|\mathbf{p}|} \sqrt{\frac{\alpha}{\pi}}\left[\exp \left(-\alpha \mathbf{p}_{\min }^2\right)-\exp \left(-\alpha \mathbf{p}_{\max }^2\right)\right],
\end{equation}
where $m_n$ is neutron mass and $\alpha=3 /\left[4\left\langle\mathbf{p}^2\right\rangle(A-2) /(A-1)\right]$ \cite{kula2006}. $\mathbf{p}_{\text {min}}$ and $\mathbf{p}_{\text {max}}$ in Eq. (\ref{35}) are the lower and upper limits of the center-of-mass momentum
\begin{subequations}
	\begin{gather}
		\mathbf{p}_{\text {min}}^2=\left\{\frac{A-2}{A-1}|\mathbf{p}|-\sqrt{2 m_n \frac{A-2}{A-1}\left[E-E^{(2)}\right]}\right\}^2, \\
		\mathbf{p}_{\max}^2=\left\{\frac{A-2}{A-1}|\mathbf{p}|+\sqrt{2 m_n \frac{A-2}{A-1}\left[E-E^{(2)}\right]}\right\}^2,
	\end{gather}
	\label{36}%
\end{subequations}
where $E^{(2)}$ is two-nucleon separation energy. In order to calculate the momentum distribution $n_{\text {corr}}(\mathbf{p})$ in Eq. (\ref{35}), we adopt the light-front dynamics (LFD) method where the one-boson-exchange model is applied to the nucleon-nucleon interaction and the parameters are taken from the Bonn potential \cite{carbonell1995relativistic}. The LFD method calculates the correlation part $n_{\text {corr}}(\mathbf{p})$ by empirically rescaling the high-momentum component of the momentum distribution of the deuteron:
\begin{equation}\label{38}
		n_{\text {corr}}(\mathbf{p})=N_\tau \mathcal{N} C_A\left[n_2(\mathbf{p})+n_5(\mathbf{p})\right] \text {, }
\end{equation}
where the two components $n_2(\mathbf{p})$ and $n_5(\mathbf{p})$ are deduced from the LFD wave functions \cite{wang2021nucleon}. Within the LFD framework, the $n_2$ originates primarily from tensor force interactions and exhibits dominance in the range $1.5 < \mathbf{p} < 3.0 {\rm \, fm^{-1}}$, while $n_5$ arises predominantly from $\pi$-meson exchange and mainly contribute at higher momentum $\mathbf{p} > 2.5 {\rm \, fm^{-1}}$. The scaling factor $C_A$ in Eq. (\ref{38}) is the ratio of high-momentum components between deuteron and target nuclei. $N_\tau$ is the normalization coefficient. The detailed formula for calculating NMD and spectral function can be found in our previous work \cite{wang2023influence,wang2021nucleon}.

\section{Numerical results and  analysis}

In this section, NMDs and spectral functions are studied with HFB model and LFD method. Based on these analyses, the $\psi$-scaling function is constructd to investigate CCQE neutrino scattering process. Moreover, we explore the CCQE scattering cross sections using the $\psi$-scaling function, including double-differential, flux-integrated, and total cross sections.

\subsection{Momentum distributions, spectral functions, and $\psi$-scaling function}

Firstly, we present the results for the nucleon momentum distributions $n(p)$, the neutron spectral functions $S(p, E)$, and the $\psi$-scaling function $f(\psi)$ of ${ }^{12} \mathrm{C}$. The corresponding nucleon single-particle wave functions in $p$-space are computed from the axially deformed HFB model using the SLy4 parameter set. The correlation part spectral functions $S_{\text {corr}}(p, E)$ are obtained using the LFD method, and the correlation strengths are $C_A = 4.5$, as specified in Eq. (\ref{38}). When calculating the $\psi$-scaling function, we use Fermi momentum $p_F=1.1{\rm \,\mathrm{fm}^{-1}}$ \cite{barbaro2018asymmetric,don2002}.

\begin{center}
	\includegraphics[width=0.5\textwidth]{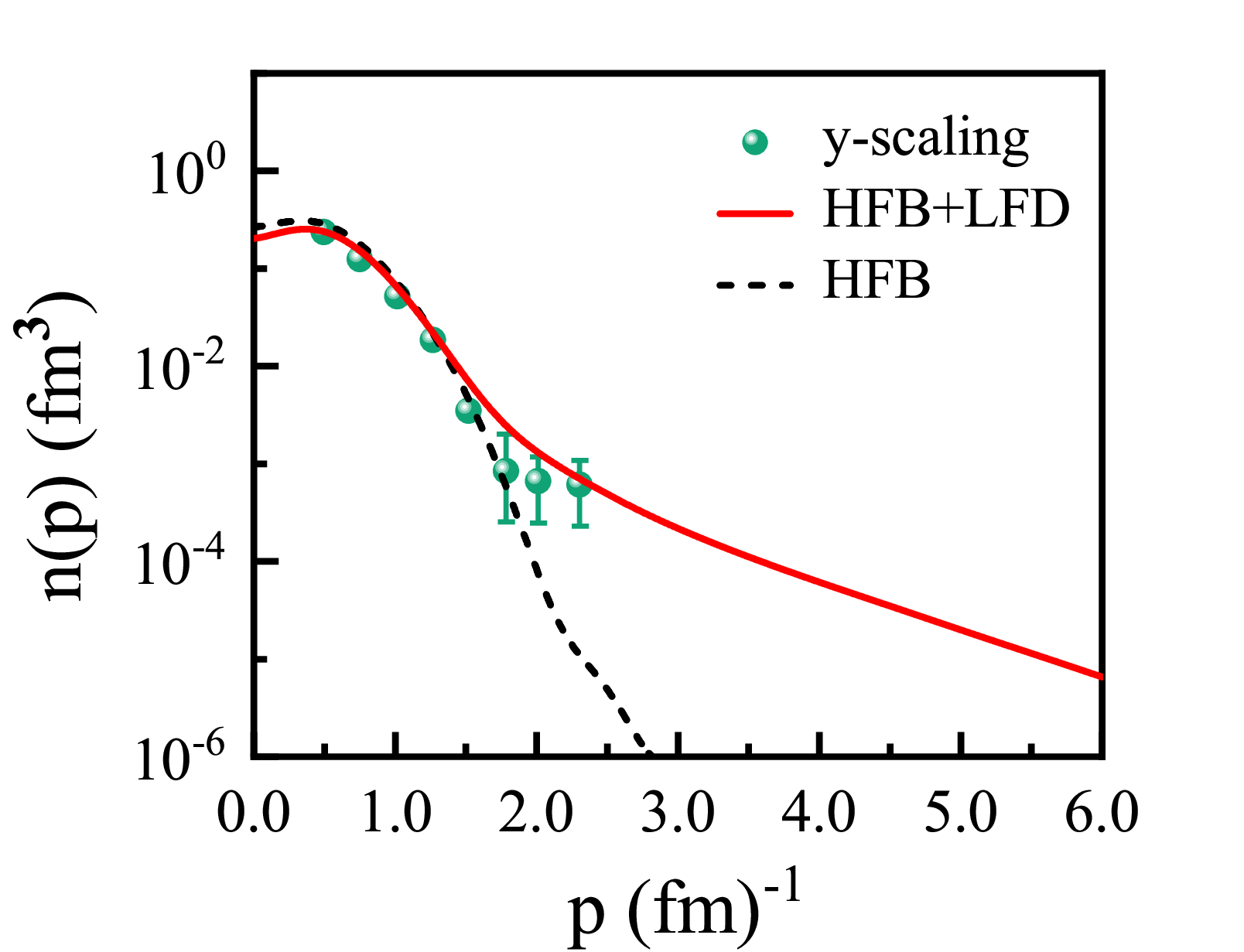}
	\figcaption{\label{fig:1}Total momentum distribution $n(p)$ of ${ }^{12} \mathrm{C}$ for the configuration $\beta = -0.1$ calculated from the deformed HFB model and the LFD method. The green balls represent the $n(p)$ obtained from $y$-scaling analyses on $\left(e, e^{\prime}\right)$ cross sections \cite{degli1991scaling}}
\end{center}

In Fig.~\ref{fig:1}, we compare the total NMDs $n(p)$ calculated using HFB and HFB+LFD methods for the configuration ($\beta = -0.1$). $\beta$ represents the quadrupole deformation of the nucleus \cite{stoitsov2013axially}. The black dashed line depicts the $n(p)$ calculated by HFB, while the red solid line represents the $n(p)$ obtained by HFB+LFD method. The NMDs extracted from the $y$ scaling analyses on $\left(e, e^{\prime}\right)$ experiments are also provided in this figure for comparison \cite{degli1991scaling}. From Fig.~\ref{fig:1}, one can see that HFB calculations provide accurate descriptions of the NMD under the Fermi momentum $p_F$. For $p>p_F$, $n(p)$ from MF model decreases rapidly and diverges from experimental data. By introducing \textit{NN}-SRC contributions with LFD method, the tail values of $n(p)$ are enhanced, yielding a $24\%$ proportion of high-momentum neutrons above the Fermi surface. This result shows excellent agreement with $y$-scaling analysis measurements. Through contrastive analysis, the LFD results in Fig.~\ref{fig:1} are also in agreement with the calculations from the realistic nucleon-nucleon interactions, such as Nijmegen-I, -II, -Reid, Argonne $v_{18}$, and Paris \textit{NN} potentials \cite{Antonov2002nucleon}.

Using the methods in Sec. 2.3, the spectral function $S(p, E)$ can be calculated by considering the $n(p)$ of both the MF and the correlation components. Fig.~\ref{fig:2} presents the logarithm of the neutron spectral functions $S(p, E)$ of ${ }^{12} \mathrm{C}$ to show the impact of \textit{NN}-SRC. It can be seen that in the regions of $E>0.05 {\rm \,GeV}$ and $p>2.0 {\rm \,\mathrm{fm}^{-1}}$, $S(p, E)$ predominantly attributes to the contributions from \textit{NN}-SRC, which has no shell structure and shows in a smooth ridge. Unlike the \textit{NN}-SRC part, one can clearly distinguish the different single-particle states in the MF  region (enclosed by a curve) in Fig.~\ref{fig:2}. It should be noted that the nuclear deformation impacts momentum distributions $n(p)$ by modifying single-particle energy levels, thereby affecting the number of nucleons participating in scattering processes, which have been addressed in our earlier work \cite{wang2023influence}.

\begin{center}
	\includegraphics[width=0.5\textwidth]{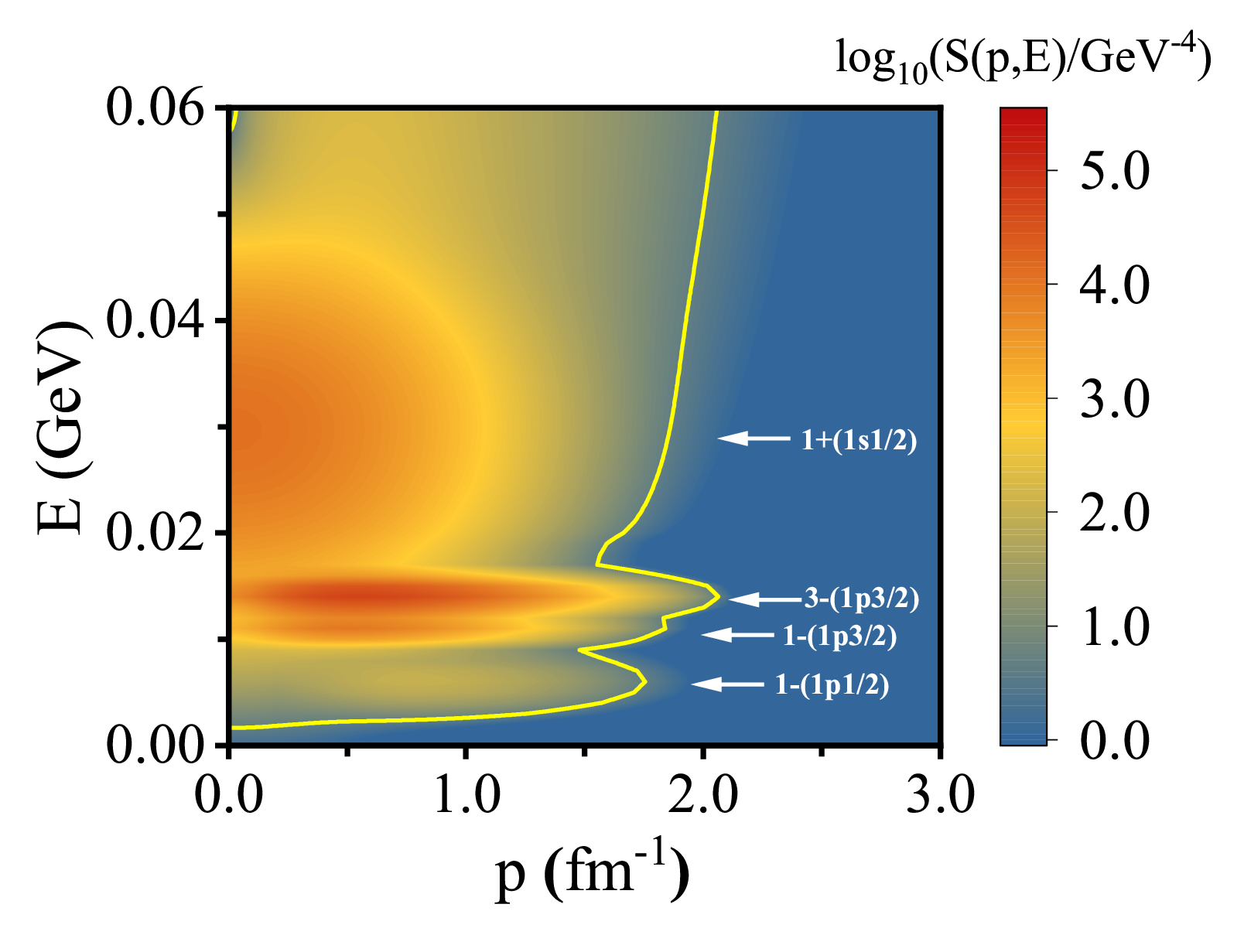}
	\figcaption{\label{fig:2}Neutron spectral functions $S(p, E)$ of ${ }^{12} \mathrm{C}$ for the configuration $\beta = -0.1$ calculated from the deformed HFB model and the LFD method. The logarithm of $S(p, E)$ is presented to highlight the \textit{NN}-SRC part, and the region enclosed by a curve describes the MF contributions.}
\end{center}

In Sec. 2.2, it is discussed that the scaling function connects the calculations of nuclear structure with the CCQE neutrino scattering process. In this part, based on the neutron spectral function $S(p, E)$ in Fig.~\ref{fig:2}, the $\psi$-scaling function $f(\psi)$ are calculated by integrating the $S(p, E)$ over the energy $E$ and momentum $p$. In Fig.~\ref{fig:3}, two $\psi$-scaling functions from HFB and HFB+LFD models are presented with the normalization $\int f(\psi) d \psi=1$. To further strengthen the credibility of our models, we also calculate and include the $f(\psi)$ from the Coherent Density Fluctuation Model (CDFM) and the experimental $f(\psi)$ extracted from electron scattering experiments \cite{donnelly1999superscaling}. As shown in this figure, two theoretical $f(\psi)$ with and without \textit{NN}-SRC can reflect the overall trend in the experimental data.

After considering \textit{NN}-SRC, the $\psi$-scaling function demonstrates different behaviors at the peak position and in the negative-$\psi$ region. At the peak position of $f(\psi)$, the result of HFB+LFD model is lower than that of the HFB model. In the tail of $f(\psi)$, the values from the HFB+LFD model are higher than those from the HFB model, especially in the region $\psi < -1.0$. This indicates that \textit{NN}-SRC mainly contribute to the low and high energy tails of $f(\psi)$. Additionally, compared to the HFB model, $f(\psi)$ with \textit{NN}-SRC exhibits asymmetry. This reflects that the strong interactions between particles induce asymmetry in the energy distribution. To enhance the credibility of our results, HFB+LFD results are compared with those from the CDFM. As can be seen from the Fig.~\ref{fig:3}, our results are consistent in behavior with the CDFM, and both of them align more closely with the experimental data.

\begin{center}
	\includegraphics[width=0.5\textwidth]{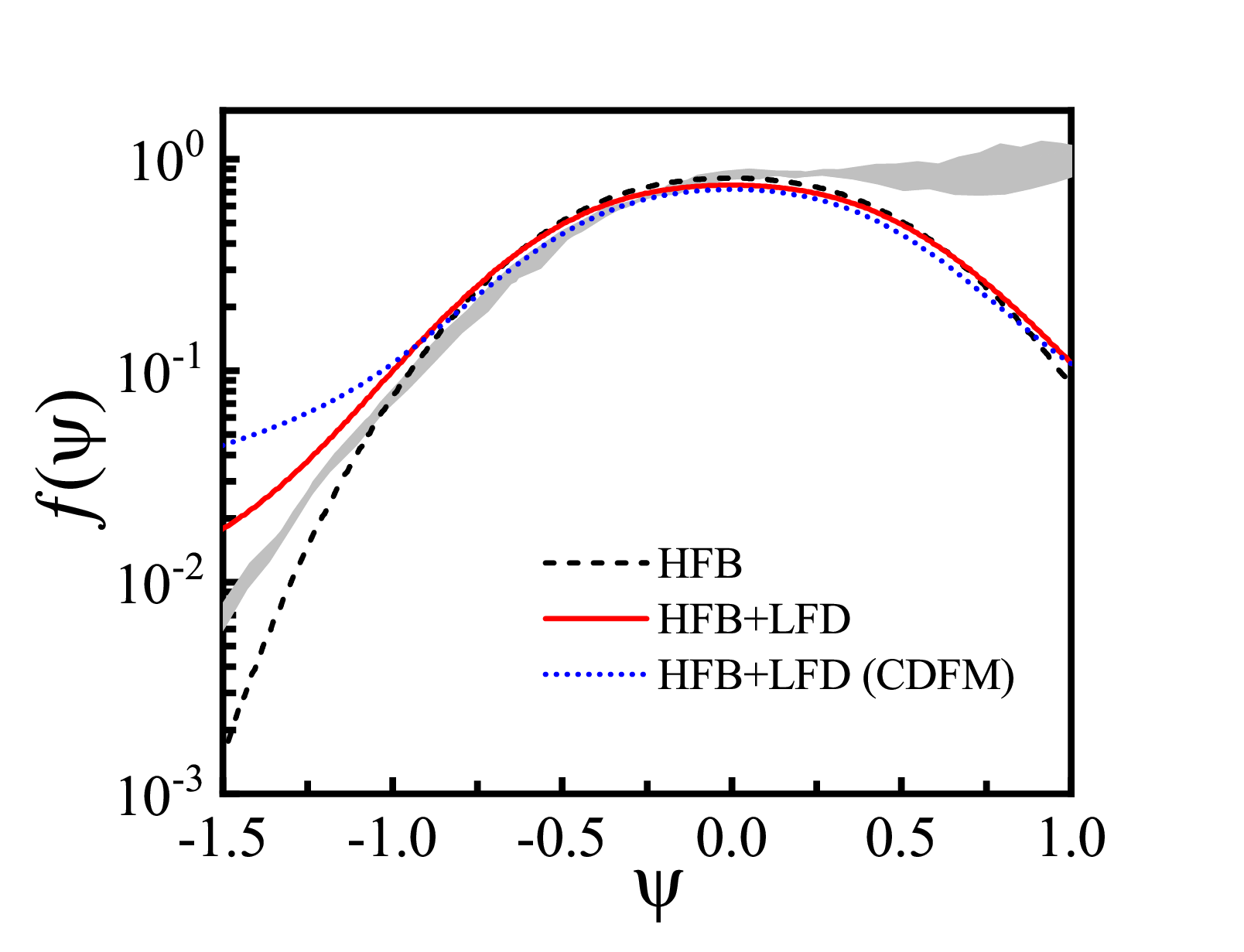}
	\figcaption{\label{fig:3}Scaling function $f(\psi)$ for ${ }^{12} \mathrm{C}$ obtained using HFB, HFB + LFD, and HFB+LFD (CDFM) models at $q=1000$ MeV/c, respectively, with the normalization $\int f(\psi) d \psi=1$. The experimental data (gray area) are from Ref. \cite{donnelly1999superscaling}.}
\end{center}

\begin{figure*}[t] 
\begin{center}
	\includegraphics[width=0.8\textwidth]{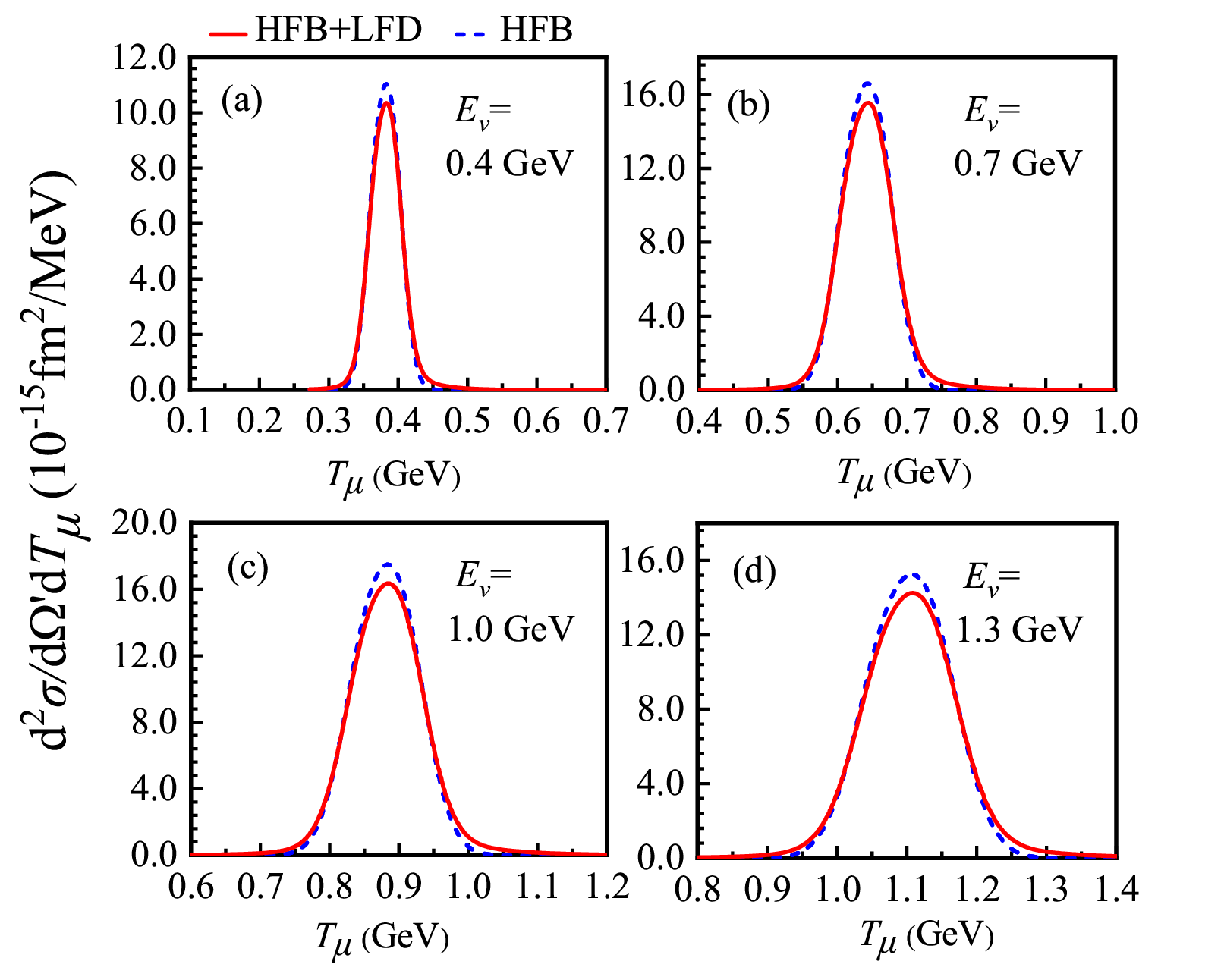}
	\figcaption{\label{fig:4}Double-differential cross sections of the reaction ${ }^{12} \mathrm{C} \left(\nu_\mu, \mu^{-}\right)$ for different incident neutrino energies $E_\nu$ at scattering angle $\theta_\mu = 30^{\circ}$.}
\end{center}
\end{figure*}

From Fig.~\ref{fig:3}, although the inclusion of \textit{NN}-SRC effects improves the behavior of $f(\psi)$, minor discrepancies persist between theoretical predictions and experimental data. This occurs because the model neglects certain complex many-body effects, including but not limited to two-particle-two-hole (2p-2h) excitations and meson-exchange currents (MEC) \cite{jea2012,gdm2016,art2016}. The 2p-2h states, describing short-range correlations as transient high-momentum nucleon pairs, introduce non-independent particle motion that significantly alters the nucleon momentum distribution, with dominant contributions in the negative-$\psi$ region \cite{vl2023}. MEC arising from non-local interactions mediated by virtual pions and other particles between nucleons, primarily induce further enhancement of the $f(\psi)$ scaling function in the negative-$\psi$ region \cite{pace2023}.

\subsection{CCQE neutrino scattering cross sections}

In this subsection, we utilize the scaling functions $f(\psi)$ to study the CCQE neutrino scattering cross sections and analyze the impact of nuclear structure on differential cross sections and total cross sections.

After obtaining the $\psi$-scaling function, the double-differential cross sections ${d^2 \sigma}/{d \Omega^{\prime} d T_\mu}$ of CCQE neutrino scattering of ${ }^{12} \mathrm{C}$ are computed by decomposing the cross sections into the product of the elementary weak scattering cross section $\sigma_0$ and the structure function $\mathcal{F}_{+}^2$ as described in Sec. 2.1. The corresponding results calculated from HFB and HFB+LFD models are shown in Fig.~\ref{fig:4}. The horizontal axis represents the kinetic energy of emitted muons, denoted as $T_\mu$. Fig.~\ref{fig:4} displays ${d^2 \sigma}/{d \Omega^{\prime} d T_\mu}$ at incident neutrino energies of $E_\nu = 0.4$, $0.7 $, $1.0 $ and $1.3 ~{\rm \,GeV}$, respectively, both at scattering angle $\theta_\mu = 30^{\circ}$.

\begin{figure*}[t] 
\begin{center}
	\includegraphics[width=0.8\textwidth]{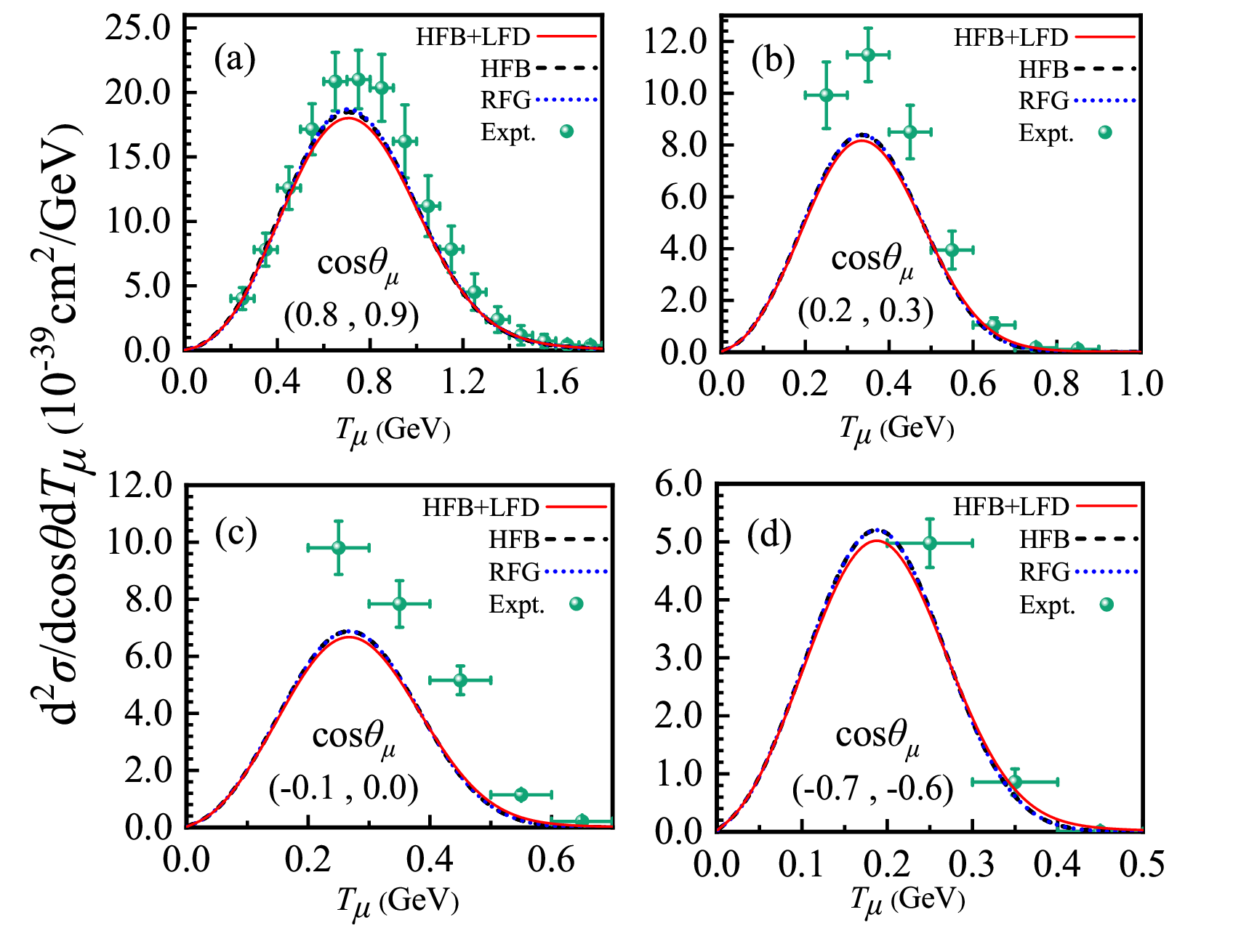}
	\figcaption{\label{fig:5}Flux-integrated double-differential cross section per target nucleon for the CCQE process on ${ }^{12} \mathrm{C}$ displayed versus the muon kinetic energy $T_\mu$ for various bins of $\cos \theta_\mu$. The corresponding scaling functions $f(\psi)$ are obtained from the RFG, HFB, and HFB+LFD approaches. (a) $0.8 \leq \cos \theta_\mu \leq 0.9$, (b) $0.2 \leq \cos \theta_\mu \leq 0.3$, (c) $-0.1 \leq \cos \theta_\mu \leq 0.0$, (d) $-0.7 \leq \cos \theta_\mu \leq -0.6$.}
\end{center}
\end{figure*}

From Fig.~\ref{fig:4}, one can observe three distinct characteristics of the cross sections. First of all, the peak of the ${d^2 \sigma}/{d \Omega^{\prime} d T_\mu}$ corresponds to the position where the scaling function $\psi=0$ in $f(\psi)$. The position of this peak represents that neutrinos are scattered by neutrons with momentum $p=0$. In addition, as $E_\nu$ increases, the location of the peak shifts towards high $T_\mu$. This is because the increase in $E_\nu$ leads to a corresponding rise in $T_\mu$. Finally, the peak position of cross section initially increases and then decreases as increasing of $E_\nu$. This behavior is linked to the interactions between neutrinos and neutrons, with a detailed discussion to follow in the total cross section.

It can also be observed in the four panels of Fig.~\ref{fig:4} that after introducing \textit{NN}-SRC, the values of ${d^2 \sigma}/{d \Omega^{\prime} d T_\mu}$ increase in the high-$T_\mu$ region. This is due to the fact that correlations lead to an increase in the number of high-$p$ neutrons in Fig.~\ref{fig:1}. Compared with the MF nucleons, the contribution of these high-$p$ nucleons becomes more pronounced at higher outgoing energies \cite{wang2023new}. Besides, the value of ${d^2 \sigma}/{d \Omega^{\prime} d T_\mu}$ at the peak position decreases in the four panels of Fig.~\ref{fig:4} after considering the contribution of \textit{NN}-SRC. For a fixed incident energy $E_\nu$ and scattering angle $\theta_\mu$, not all neutrons can participate in the CCQE scattering process and a certain portion of neutrons are precluded due to the requisite conditions of energy and momentum in Eq. (\ref{23}). Therefore, after incorporating \textit{NN}-SRC, the cross section is lower than the cross sections with only MF contributions.

The quasielastic ${ }^{12} \mathrm{C} \left(\nu_\mu, \mu^{-}\right)$ flux-integrated differential cross sections ${d^2 \sigma}/{d  \cos \theta d T_\mu}$ are analyzed using the formula from Eq. (\ref{18}). The corresponding results are displayed in Fig.~\ref{fig:5}. In this figure, $f(\psi)$ of RFG model is refer to Ref. \cite{amaro2005using}. The experimental data and the neutrino flux $\Phi\left(E_\nu\right)$ are sourced from the MiniBooNE experiment \cite{aguilar2010first}. From Fig.~\ref{fig:5}, one can see that three theoretical results can effectively reproduce the shape and the positions of the peaks at different scattering angles. Besides, the flux-integrated differential cross sections all start from $T_\mu=0$. This because the Eq. (\ref{18}) accounts for all incident neutrino energies, resulting in the outgoing muon kinetic energy start from $T_\mu=0$. Finally, the peak width of the scattering cross sections narrows as the $\cos \theta_\mu$ decreases. This narrowing is attributed to the momentum triangle relationship ${q}^2={k}^2+{k^{\prime}}^2-2{kk}^{\prime} \cos \theta_\mu$ and momentum conservation in Eq. (\ref{26}), where a smaller $\cos \theta_\mu$ results in larger momentum transfer $q$, leading to a corresponding decrease in the maximum value of $T_\mu$.

After introducing \textit{NN}-SRC, it is clearly visible in Fig.~\ref{fig:5} that, compared with the RFG and HFB models, the HFB+LFD model displays higher values and extends further in the right tail of the cross sections. This is because the other two models lack high-$p$ neutrons, with all states below the Fermi momentum $p_F$ being occupied. Besides, it can be observed that after introducing \textit{NN}-SRC, there is a slight decreasing in the peak position of flux-integrated differential cross sections.

\begin{center}
	\includegraphics[width=0.5\textwidth]{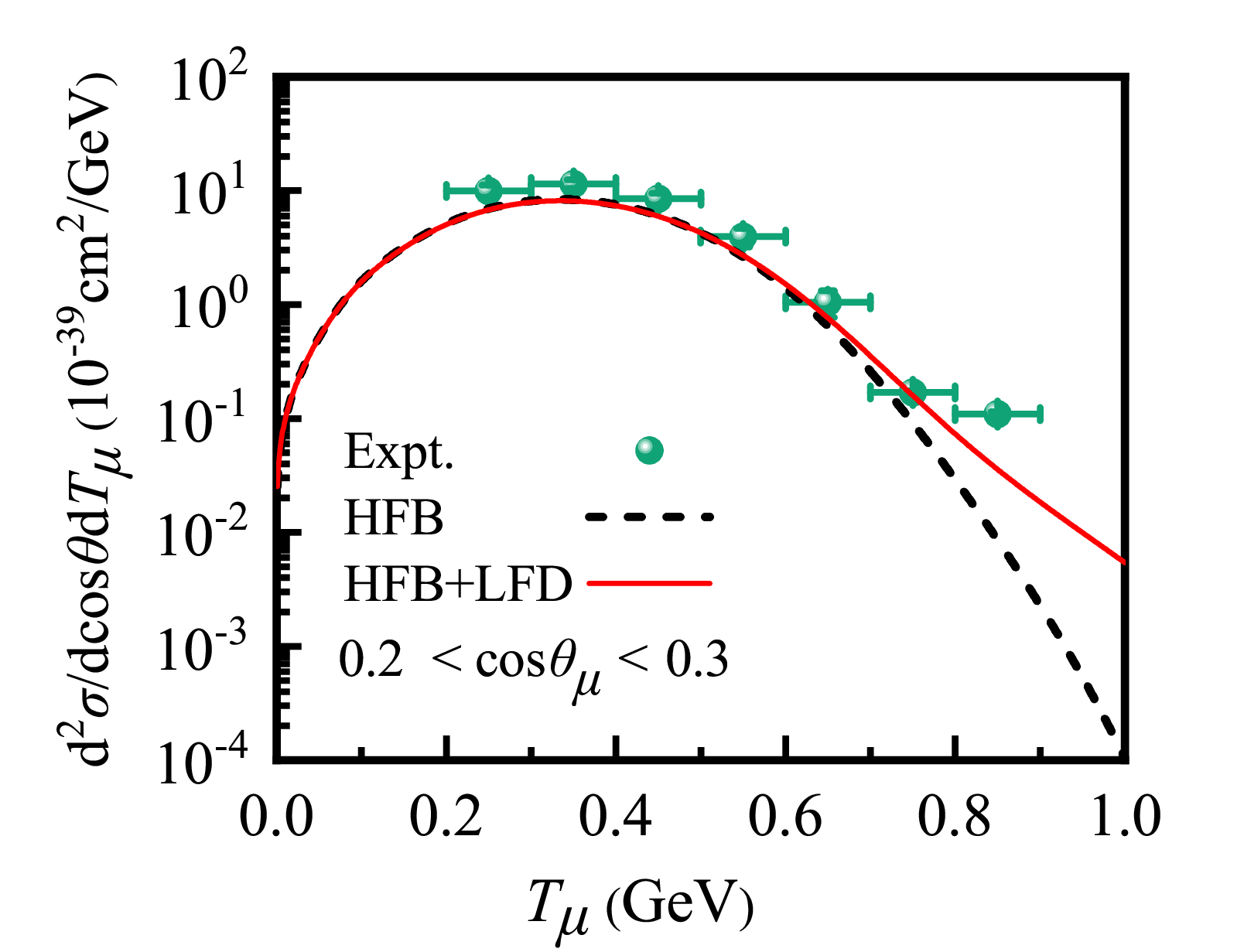}
	\figcaption{\label{fig:10}Flux-integrated double-differential cross section per target nucleon on ${ }^{12} \mathrm{C}$ versus muon kinetic energy $T_\mu$ in logarithmic coordinates, with the scattering angle range $0.2 \leq \cos \theta_\mu \leq 0.3$. In calculation process, the correlation strengths $C_A = 4.5$, corresponding to $24\%$ of neutrons above the Fermi momentum $p_F$.}
\end{center}

In order to more clearly extract information about the neutron momentum distribution from the scattering cross section and analyze the influence of \textit{NN}-SRC on neutrino scattering, Fig.~\ref{fig:10} presents the quasielastic ${ }^{12} \mathrm{C} \left(\nu_\mu, \mu^{-}\right)$ flux-integrated differential cross sections ${d^2 \sigma}/{d  \cos \theta d T_\mu}$ in logarithmic coordinates, with the scattering angle range of $0.2 \leq \cos \theta_\mu \leq 0.3$. As shown in Fig.~\ref{fig:10}, the HFB+LFD model, which incorporates \textit{NN}-SRC effects, yields larger values with an extended tail compared to the results of HFB model. This behavior can be explained through the scaling function $f(\psi)$ in Fig.~\ref{fig:3}. By combining the definition of $\psi$ in Eq. (\ref{22}) and the energy conservation relation $T_\mu=E_\nu-\omega$, the negative-$\psi$ region in Fig.~\ref{fig:3} corresponds to the high-$T_\mu$ region of the scattering cross section. Because \textit{NN}-SRC effects predominantly influence the negative-$\psi$ region, therefore, the enhanced cross section observed in the high-$T_\mu$ region of Fig.~\ref{fig:10} is directly attributable to \textit{NN}-SRC effects.

In Fig.~\ref{fig:10}, the HFB+LFD theoretical calculations incorporating \textit{NN}-SRC effects align well with experimental values, exhibiting a upward trend at the right tail of the scattering cross section ($T_\mu > 0.6 {\rm \,GeV}$). Furthermore, In Fig.~\ref{fig:10}, comparison between theoretical results and experimental data reveals that the \textit{NN}-SRC strength is constrained to $C_A = 4.5$ in Eq. (\ref{38}) for ${ }^{12} \mathrm{C}$. This exhibits an upward trend in the high-$T_\mu$ region of the scattering cross section and indicates that correlated neutrons account for approximately $24\%$ of the total neutron population. In this analysis, the contribution of \textit{NN}-SRC effects to high-momentum neutrons in ${ }^{12} \mathrm{C}$ consistent with the \textit{ab initio} calculation.

We continue to analyze the trend of scattering cross sections ${d \sigma}/{d  \cos \theta_\mu}$ with respect to variations in outgoing muon angle $\theta_\mu$. Based on Eq. (\ref{20}), the cross sections are represented as a function of scattering angle $\theta_\mu$ by integrating over the incident neutrino energy $E_\nu$ and the outgoing muon kinetic energy $T_\mu$. In Fig.~\ref{fig:6}, cross sections ${d \sigma}/{d  \cos \theta_\mu}$ from three models are compared to analyze the impact of neutron momentum distributions on the cross sections. From this figure, one can see that scattering primarily occurs at small angles and is sensitive to the changes in $\theta_\mu$. Additionally, the results of the three models exhibit minimal differences in Fig.~\ref{fig:6}. This is because the scattering involves the same neutron number $\mathcal{N}$ but different NMDs. It is discernible that, compared to the NMDs, ${d \sigma}/{d  \cos \theta_\mu}$ are more responsive to the neutron number $\mathcal{N}$ involved in the CCQE process.

\begin{center}
	\includegraphics[width=0.5\textwidth]{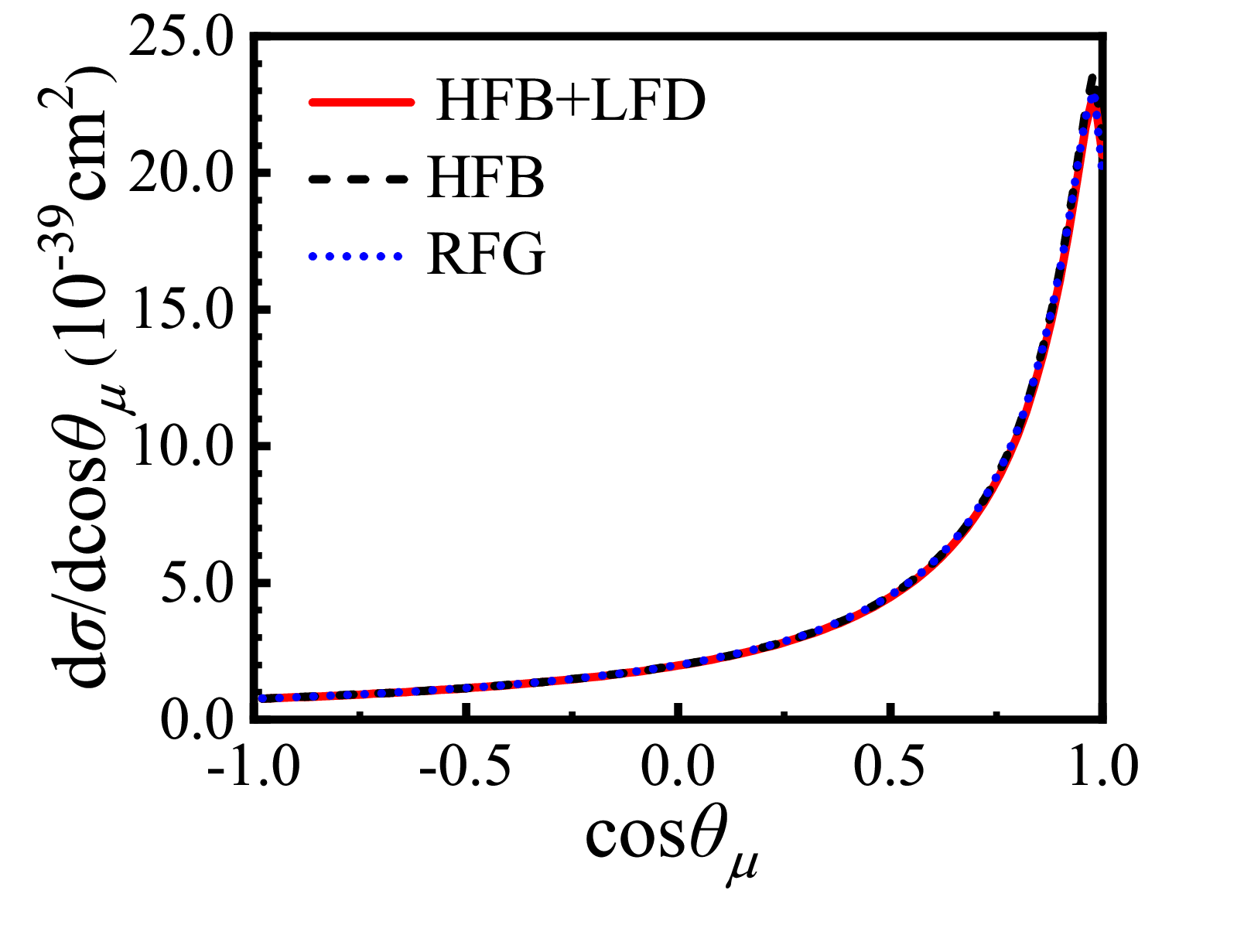}
	\figcaption{\label{fig:6}Flux-averaged CCQE $\nu_\mu$-${ }^{12} \mathrm{C}$ neutrino scattering differential cross section per nucleon as a function of the muon scattering angle $\theta_\mu$.}
\end{center}

At the end of this section, the total cross sections $\sigma_T$ of CCQE neutrino scattering with Eq. (\ref{21}) are investigated and the results of $\sigma_T$ are presented in Fig.~\ref{fig:7}. In this figure, the blue line, the black line, and the red line represent the total cross section calculated using the RFG, HFB, and HFB+LFD models, respectively. For comparison, the experimental data from the MiniBooNE and NOMAD are also provied in this figure. As shown in Fig.~\ref{fig:7}, the shapes of the total theoretical cross sections $\sigma_T$ predicted by three theoretical models agree well with the experimental data, which demonstrates the effectiveness of scaling theory and the reliability of spectral function theory in Subsec. 2.

There are three features displayed in Fig.~\ref{fig:7}. Firstly, the total cross section $\sigma_T$ begins at reconstructed neutrino energy $E_\nu = 0.105 {\rm \,GeV}$, which is due to the minimum energy of the incident neutrinos must be sufficient to emit a muon with $m_\mu^{\prime} = 0.105 {\rm \,GeV}$. Secondly, in the region $0.105 {\rm \,GeV}<E_\nu < 1 {\rm \,GeV}$, the interaction between neutrinos and neutrons strengths progressively, leading to a continuous increase in the value of the $\sigma_T$. This can also explain the rise of the double-differential cross section ${d^2 \sigma}/{d \Omega^{\prime} d T_\mu}$ with the increase of $E_\nu$ in Fig.~\ref{fig:4}. Finally, it can also observe that in the region $E_\nu > 1 {\rm \,GeV}$, the $\sigma_T$ essentially stabilizes with the increase of the $E_\nu$. This indicates that the number of nucleons involved in CCQE neutrino scattering reaches saturation.

\begin{center}
	\includegraphics[width=0.5\textwidth]{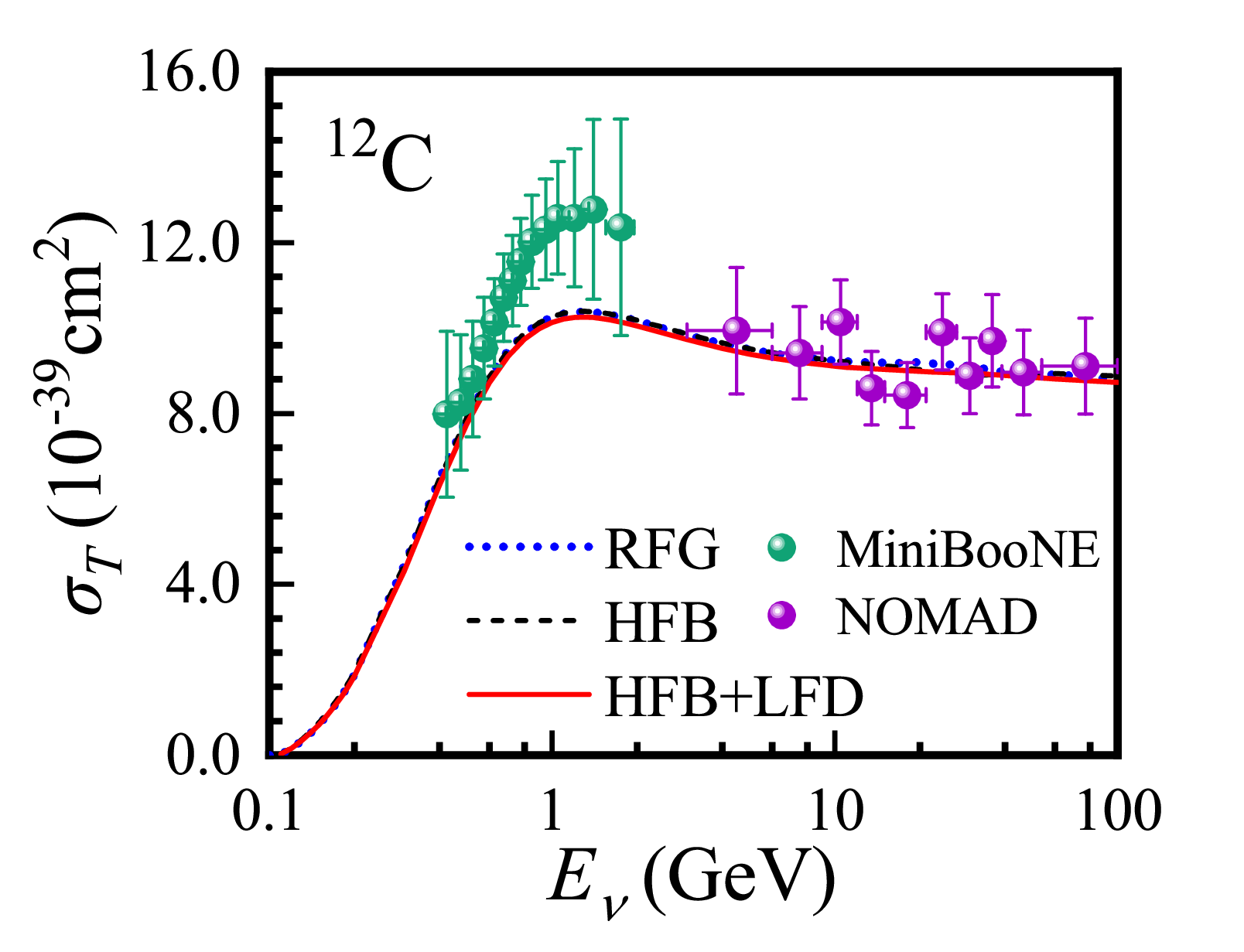}
	\figcaption{\label{fig:7}CCQE $\nu_\mu$-${ }^{12} \mathrm{C}$ neutrino scattering total cross sections per nucleon are displayed versus reconstructed neutrino energy $E_\nu$, evaluated using the RFG, HFB, and HFB + LFD approaches. The experimental data are from MiniBooNE \cite{aguilar2010first} and NOMAD \cite{lyubushkin2009study} experiments.}
\end{center}

We continue to compare and analyze the results from the three theoretical models. It is worth noting that after including \textit{NN}-SRC, the difference in the total $\sigma_T$ is minor between the HFB+LFD and the HFB models. This is also because the scattering cross sections are more sensitive to the neutron number rather than to the neutron momentum distribution. Therefore, after integrating over $\theta_\mu$ and $T_\mu$ for different models in Fig.~\ref{fig:4}, the differences become negligible in total cross sections. If we studying \textit{NN}-SRC effects through the CCQE neutrino scattering, the flux-integrated cross sections ${d^2 \sigma}/{d  \cos \theta d T_\mu}$ in Fig.~\ref{fig:5} may present a more advantageous approach than the total cross section in Fig.~\ref{fig:7}.

The results from both Figs.~\ref{fig:5} and~\ref{fig:7} also show a reduction in flux-integrated and total cross sections for both the HFB and HFB+LFD compared to experimental values. This discrepancy primarily arises from the theoretical calculations not accounting for two-nucleon knockout processes \cite{aguilar2010measurement} and 2p-2h effect \cite{sobc2025}. In the CCQE process, the occurrence of two-nucleon knockout increases the overall probability of experimental scattering events. When neutrinos interact with groups of nucleons, such as proton-neutron pairs, the probability of neutrino-nucleus interactions is enhanced, resulting in an increased experimentally measured cross section. This phenomenon results in theoretical results that are lower than the experimental data.

\section{Summary}

In this work, we develop the theoretical model of CCQE neutrino scattering which includes two primary components: the single nucleon scattering cross section $\sigma_0$ derived from the nucleon form factors, and the scaling function $f(\psi)$ come from the sophisticated nuclear structure model. Using the constructed scattering model, we analyze the behavior of CCQE neutrino scattering cross sections at different scattering angles $\theta_\mu$ and incident neutrino energies $E_\nu$  to investigate the contributions of the MF and correlation nucleons.
	
	The findings are summarized as follows. From flux-averaged differential cross sections, one can see that the theoretical results can effectively reproduce the shape and the positions of the peaks at different scattering angles. The neutron momentum distribution can be extracted from the cross sections, elucidating the impact of \textit{NN}-SRC effects on neutrino-nucleus scattering. For the total cross section $\sigma_T$, the behavior of scattering cross sections are analyzed across different reconstructed neutrino energy ranges. The starting point of the scattering cross sections occur at $E_\nu = 0.105 {\rm \,GeV}$. With the increase of $E_\nu$, a turning point appears at $E_\nu = 1 {\rm \,GeV}$. In the region $E_\nu > 1 {\rm \,GeV}$, the total scattering cross section remains stable. This suggests that the number of nucleons participating in CCQE neutrino scattering achieves saturation.

The CCQE models in this paper not only examine the nuclear structure models but also can be used to study the momentum distribution of neutrons. The studies in this paper enhance our understanding of nuclear structure and provide essential constraints for the analysis of signals and backgrounds in future neutrino oscillation experiments.

\vspace{3mm}

\vspace{-1mm}
\centerline{\rule{80mm}{0.1pt}}

\end{multicols}

\clearpage

\end{document}